\documentclass[fleqn,twoside]{article}
\usepackage{espcrc2}
\usepackage{epsf}
\usepackage[figuresright]{rotating}

\newcommand{\AmS}{{\protect\the\textfont2
  A\kern-.1667em\lower.5ex\hbox{M}\kern-.125emS}}

\newcommand{\Msun}{M$_{\odot}$}
\newcommand{\HI}{H\kern0.1em{\scriptsize I}~}
\newcommand{\mHI}{\rm H\kern0.1em{\scriptsize I}~}

\def\aj{{AJ}}
\def\araa{{ARA\&A}}
\def\apj{{ApJ}}
\def\apjl{{ApJ}}

\def\apss{{Ap\&SS}}
\def\aap{{A\&A}}

\def\mnras{{MNRAS}}

\newcommand{\simgt}{\lower.5ex\hbox{$\;\buildrel>\over\sim\;$}}
\newcommand{\simlt}{\lower.5ex\hbox{$\;\buildrel<\over\sim\;$}}

\newcommand{\nusnr}{$\nu_{\rm SNR}$}

\newcommand{\sbs}{SBS\,0335$-$052}

\title{From gas to galaxies}

\author{J.M. van der Hulst \address[KI]{Kapteyn Institute, Postbus 800,
 NL-9700 AV Groningen, the Netherlands}, 
  E.M. Sadler \address[SyU]{School of Physics, Sydney University, Sydney, Australia},
  C.A. Jackson \address[ATNF]{ATNF, CSIRO, Epping, Australia},
  L.K. Hunt \address[IRF]{Instituto di Radioastronomia-Sezione Firenze, Florence, Italy},
  M. Verheijen \addressmark[KI],
  J.H. van Gorkom \address[NY]{Columbia University, New York, U.S.A.}
  }
\begin{document}
\begin{abstract}

  The unsurpassed sensitivity and resolution of the Square Kilometer
  Array (SKA) will make it possible for the first time to probe the
  continuum emission of normal star forming galaxies out to the edges
  of the universe. This opens the possibility for routinely using the
  radio continuum emission from galaxies for cosmological research as
  it offers an independent probe of the evolution of the star
  formation density in the universe. In addition it offers the
  possibility to detect the first star forming objects and massive
  black holes.
  
  In deep surveys SKA will be able to detect \HI in emission out to
  redshifts of $z \approx 2.5$ and hence be able to trace the
  conversion of gas into stars over an era where considerable
  evolution is taking place. Such surveys will be able to uniquely
  determine the respective importance of merging and accreting gas
  flows for galaxy formation over this redshift range (i.e. out to
  when the universe was only one third its present age). It is obvious
  that only SKA will able to see literally where and how gas is turned
  into stars.
  
  These and other aspects of SKA imaging of galaxies will be discussed.

\vspace{1pc}
\end{abstract}

\maketitle

\section{Introduction}

More and more evidence supports the paradigm that galaxies form
hierarchicallly, i.e. that present day galaxies have been
built up over cosmological time scales by the coalescence of many
smaller, less massive galactic systems 
\cite{baugh1,klypin1,moore1,steinmetz1,cole1,abadi1}.  
For a particular galaxy this process is described
by the so-called merger tree: the accumulation of dark matter mass
over time into a final halo. These merger trees are extracted from
numerical comological simulations and only describe the merging of
dark matter haloes in a robust manner.

Many properties of galaxy populations including current star-formation 
rates, luminosity functions, the Tully-Fisher relation,
morphology segregation, and the variation of star-formation histories
with environment are explained by semi-analytical, hierarchical models. 
However, the role and fate of the gas and stars in this merging process is 
not well understood due to an incomplete understanding of the relevant
physics, the simplicity of the hydrodynamical codes and the lack of
numerical resolution.  The behaviour of the baryons in these merging
dark matter haloes is often described with semi-analytical techniques
\cite{kauffmann,benson1,cole1,cole2} (see also Baugh et al. (this
volume)).  In this scheme, the physics of galaxy formation is encoded
in a set of rules that are either physically motivated or have been
designed to reproduce the results of numerical simulations e.g. the
consequences of galaxy mergers. A merger tree (the ``family history'')
of a dark matter halo is either generated using Monte Carlo techniques
or is extracted from an N-body simulation. The uppermost branches of
the tree are filled with gas that is shock heated to the virial
temperature of these progenitor halos, and a set of rules is
followed to describe the subsequent evolution of this gas. Apart from
the issue of properly modelling the fate of the baryons, it is also
true that the typical shape of a galaxy merger tree has not been
confirmed robustly from observations.  Observationally, the galaxy
mass functions and the merger rates as a function of redshift are
largely unknown.

In this picture there still are several key questions that need to be
addressed. These are:
\begin{itemize}
\item [-] When and how quickly did the gas settle into the dark matter
  potential wells?
\item [-] Where, when and how did the gas get converted into stars?
\item [-] When and how did the merging process of small galaxies develop?
\item [-] What is the relative importance of gas accretion versus galaxy 
  merging?
\item [-] Are galaxies still (trans)forming today?
\end{itemize}
In order to address these questions it is necessary to have a census
of galaxies and their properties over a large range of environments
and as large a possible range of cosmic time or redshift. While
the stellar content of galaxies can be probed using optical and NIR
imaging and spectroscopy using the best facilities on the ground and
in space, the only way to probe the cold gas content of galaxies is to 
use a radio telescope, either at very short wavelengths to measure
molecular line emission (see also Blain et al. this volume 
on molecular lines), 
or at long wavelengths to measure the \HI
emission using the 21-cm line. To do this out to redshifts of $z = 3$,
since when considerable evolution may have taken place
\cite{storrielombardie1,giavalisco1}, 
requires a radio telescope tenfold larger (in collecting area)
than any of the existing facilities. The SKA is such a
telescope and the purpose of this chapter is to review the current
state of affairs, from both the observational and theoretical points 
of view,
and to outline how the SKA will be instrumental in answering the above
questions to which we do not yet have firm answers.

Equally as important, SKA can probe the star formation rates in galaxies
using the radio continuum emission. The radio continuum emission
provides an estimate of the star
formation density with redshift independent of other tracers such as
optical or ultraviolet light which is affected by extinction. 
The caveat is that the
relation between the non-thermal radio emission and the star formation
rate is assumed not to vary significantly with cosmic time.

The framework for the discussion in this chapter will be the
hierarchical structure formation models of Baugh et al. (this volume), 
who use the technique of semi-analytic modelling to
predict the evolution of the HI mass function and the number of star
forming radio galaxies, both signposts of evolution to be probed with
SKA. This follows a widely accepted idea of how galaxies form, based
on the success of the Cold Dark Matter ($\Lambda$CDM) simulations.

Although $\Lambda$CDM has proven to be very succesful in reproducing
the WMAP results and resulting large scale structure, SKA will observe
the baryonic component of the universe directly. Predictions of what
we should observe are still wide open, as recent hydrodynamical
simulations indicate that some of the basic assumptions of the
semi-analytic models, i.e. that the gas gets heated quickly to the
virial temperature, may not be correct.  In the semi-analytic
modelling the assumption is that the gas gets immediately shock heated
to the virial temperature as it flows into the dark matter potential
wells. Recently Binney \cite{binney1}, Katz et al.~\cite{katz1} and
Murali at al.~\cite{murali1} have pointed out that hydrodynamical
simulations suggest that most of the gas is not heated to the virial
temperature but accretes slowly in a cool phase from the surrounding
large scale structure filaments. The consequence is that galaxies grow
through accretion of gas rather than merging of dark matter halos.  
At redshifts beyond $z
\approx 2$ this cool mode of accretion appears to dominate over
merging as the main process for galaxy formation. Because these two
scenarios differ in the evolutionary scenario of the gas, SKA is the
ultimate telescope for distinguishing these from one another.

We will first discuss the properties of radio continuum emission from
galaxies and in this context the first luminous objects in the universe. 
Then we will discuss the subsequent evolution of galaxies following
the theoretical scenarios, and finally connect this with the evolution
of galaxies which is still ongoing and observable at the present time
in the local universe.


%
%




\section{The radio continuum emission from star--forming galaxies}

\noindent
When and how the first episodes of star formation took place remains 
a key question of modern cosmology.  Much effort has been devoted to 
measuring the integrated cosmic star formation rate (SFR) of the 
universe as a function of redshift or look-back time, initially at 
ultraviolet wavelengths \cite{madau}, and more recently 
in the millimetre and radio regimes \cite{blain99,haarsma}.
Unlike ultraviolet or optical light, radio continuum emission  
is unaffected by dust, which eliminates the need for uncertain 
extinction corrections.  SKA will detect the radio continuum 
emission from very large numbers of AGN and star--forming galaxies 
at redshifts $z>3$.
However, we caution that the conversion between a measured radio 
continuum flux and a derived a global SFR requires the adoption of 
canonical scaling relations \cite{condon2} which remain valid 
out to at least $z\sim$1 \cite{garrett02,appleton1} but, as we discuss 
below, may not be directly applicable to primordial star-forming systems 
at $z>3$. 

In the models considered here, the radio continuum emission from 
star--forming galaxies is computed using the spectro-photometric 
code {\tt GRASIL} written by Silva et al.\cite{silva}. This code takes 
star--formation histories generated by {\tt GALFORM}, including 
bursts of star formation, and produces a full spectral energy 
distribution, ranging from the far-UV to the radio. The radio 
emission includes both synchroton radiation by electrons accelerated 
in supernova remnants and free--free emission from HII regions.
The plots presented in  this section are for a model very similar 
to that of Baugh et al. (this volume). Note that beyond redshifts of
$z > 6$ inverse compton losses against the cosmic microwave background
will begin to affect the synchrotron emission. The non-relativistic, 
thermal emission is not affected \cite{carilli99} so that radio continuum
emission still is a very good tracer of star formation at high redshifts,
albeit in another regime.

\begin{figure}[htb]
\vspace*{8cm}
\includegraphics{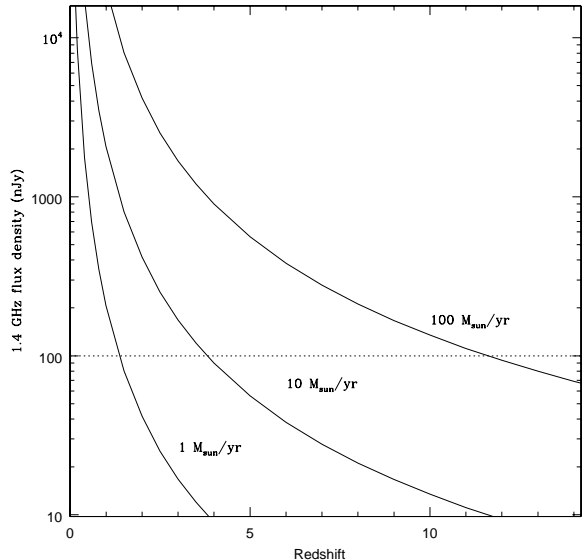}
\caption{Predicted 1.4\,GHz flux densities for star--forming galaxies 
(with SFRs of 1, 10 and 100\,M$_\odot$/yr) at different redshifts.  
The horizontal line at 100\,nJy represents the nominal SKA detection 
limit for an 8\,hr integration. 
}
\label{fig:sf_flux}
\end{figure}

\noindent
\subsection{The first luminous objects (z$>$5) } 

Little is currently known about the properties of star--forming 
objects at redshifts above z$\sim5$ but the next few years are 
likely to see rapid progress in this area. 

Several recent results suggest substantial star formation at high
redshift \cite{beck}. 
Studies of the stellar `fossil record' of nearby galaxies
for example
(e.g. \cite{heavens}) imply that the star--formation rate in the
most massive galaxies increases with redshift and peaks at z=3 or
higher. The recent detection of QSOs at redshifts with z$>$6 
\cite{fan}, in which the gas appears to have solar or above-solar
metallicity \cite{pentericci} suggests that the massive first
stars in the Universe had already formed by z=8 or even earlier.
Observations with the HST and ground--based optical/IR telescopes have
begun to provide direct measurements of the number of star--forming
galaxies at $z\sim$\, 5--6 \cite{bouwens,yan,hu,stanway}. These studies 
find typical surface densities of $\sim100$\,deg$^{-2}$ for star--forming
galaxies at $z\sim6$.  So it appears that the global star--formation rate
at $z\sim6$ is roughly similar to that at $z\sim$\,3--4, though with
substantial uncertainities because the distribution of high--redshift
galaxies appears to be clustered on scales comparable to the diameters
of the individual fields studied to date \cite{hu}.

At even higher redshift, the detections of individual
(gravitationally--lensed) galaxies at $z\sim7$ \cite{kneib}
and $z=10$ \cite{pello} have recently been reported, but the
density of star--forming galaxies at $z>7$ is essentially unknown at
present.  

SKA will be a sensitive probe of high-redshift star formation through
the detection of radio continuum emission from star--forming regions
and the radio afterglows of gamma--ray bursts (GRBs), many of which
are now known to arise from the supernova explosions of massive stars.

Models for dwarf galaxies forming in the vicinity of high--redshift
QSOs imply likely star--formation rates of 30--60 M$\odot$/yr for
these objects \cite{natarajan}.  Deep continuum surveys at
1.4\,GHz with SKA will reach detection limits of 100\,nJy in 8\,hr,
and as low as 10\,nJy in longer integrations of $\sim$1000\,hr 
\cite{taylorbraun}.  This should allow objects with star--formation rates
as low as 10 M$\odot$/yr to be detected at redshifts up to z$\sim$4
(8\,hr) and z$\sim$11 (in 1000\,hr) respectively (see Fig.~\ref{fig:sf_flux}).
Individual globular clusters forming at high redshift may also be
detectable by SKA, though there are currently few available
predictions for the star--formation rates in such objects.  


\noindent
\subsection{What will SKA see? } 
SKA will be able to detect very faint (10--100\,nJy) radio continuum 
sources, i.e. 2--3 orders of magnitude fainter than those seen by 
present--day telescopes.  It is therefore important to keep in mind 
that simulations of the likely surface density, angular size and 
redshift distribution of these faint sources require extrapolations 
from our current knowledge (as well as assumptions about the cosmic 
evolution of each population). Nevertheless, the work done so far 
(e.g. \cite{hopkins,garrett02,windhorst}) gives us 
some valuable ideas about what to expect. At 100\,nJy, the surface 
density of continuum sources will be very high 
($\sim$200,000\,deg$^{-2}$), with ``normal'', starburst and active 
galaxies all contributing over a wide range in redshift.  Simulations 
by Hopkins et al. \cite{hopkins} imply that very high--redshift ($z>5$) 
AGN and starburst galaxies will have surface densities of 
$\sim$1,000\,deg$^{-2}$ and $\sim$7,000\,deg$^{-2}$ respectively 
in surveys with a detection limit of 100\,nJy at 1.4\,GHz.

\noindent
\subsection{Distinguishing starbursts from AGN}

Although star--forming galaxies dominate the radio source counts at
flux densities below about 1\,mJy, active galaxies still account for a
significant fraction of the population at levels of a few $\mu$\,Jy
\cite{Hammer}.  The problem of distinguishing star--forming galaxies
from AGN is discussed by Garrett et al.  \cite{garrett01}. Several
parameters measurable by the SKA itself (including brightness
temperature, morphology, time variability and possibly polarization)
can be used, and Garrett et al.\cite{garrett01} show that radio continuum
observations with high spatial resolution ($\sim25$\,mas) can reliably
distinguish AGN from star forming galaxies at faint flux levels. This
would require the SKA to have good sensitivity on baselines of at
least 3000\,km.  High spatial resolution is also required because of
the likely source density at flux levels of $\sim$10\,nJy, which may
be as high as 3\,arcsec$^{-2}$ \cite{taylorbraun}.\\


\subsection{Probing high--redshift star formation through GRBs}

\noindent
Because of their transient nature, which makes them easy to recognize,
and because they appear to be associated with the supernova explosions
of young, very massive stars, GRBs may be one of the most direct
tracers of star formation at very high redshifts.  SKA, with its wide
field of view, should be able to detect the radio afterglows of many
`orphan' GRB events in which the jet is misaligned with the observer's line
of sight, and thus remains undetected in gamma--rays (e.g. Frail et al 
\cite{frail1}). 
Bromm \& Loeb \cite{bromm} predict that at least 50\% of GRBs
originate at redshift $z>5$, and Berger et al.\cite{berger} find that GRB
host galaxies have typical star--formation rates of at least
100\,M$_\odot$/yr.  In the next decade, therefore, GRBs are likely to
become increasingly common probes of high--redshift star formation up
to (and perhaps even within) the epoch of reionization.  


\subsection{Redshift measurements and estimates}

Although SKA will be able to detect many millions of star--forming
galaxies over a very wide range in redshift, recognizing the
highest--redshift objects will not be straightforward.  For most
star--forming galaxies with z$<$3, redshifts should be available from
the large--area HI surveys which SKA itself will carry out.
Alternatives at higher redshift include: \\ 
(i) HI absorption lines (\cite{carilli}; see also Kanekar and Briggs,
this volume). \\ 
(ii) CO emission--line redshifts, using either the low
order transitions observed
with SKA, or higher--order transitions observed with ALMA. 
(see Blain et al. this volume) \\ 
(iii) Radio 'photometric redshifts' based on the spectral slope between
1.4\,GHz and 350 GHz \cite{yun}.  This would
require matching high-frequency observations from ALMA. \\ 
(iv) Photometric redshift estimates from wide-field multicolour optical/IR
images, or spectroscopic measurements in the optical/IR for small
samples of particular interest.  \\



\section{How different is high--redshift star formation?}

In this section, we examine the assumptions underlying the use
of the radio continuum to measure SFR, and assess their validity at
the high redshifts where SKA will be able to make a unique contribution.
We also outline new simulations which will enable us to better
interpret the planned deep surveys.

\begin{figure}[htb]
\vspace*{7cm}

\includegraphics{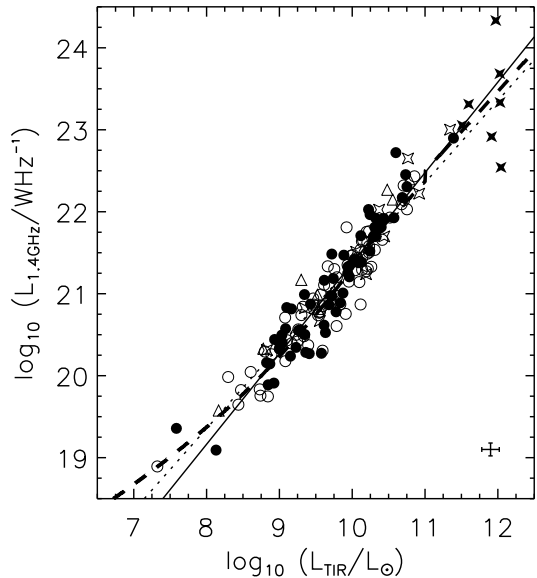}

\caption{1.4\,GHz radio luminosity (log) plotted
against total far--infrared luminosity (taken from Bell 2003 \cite{bell}), 
showing the curvature at low luminosities. Normal star--forming galaxies
are plotted as circles, intensely star--forming galaxies as stars (filled
stars are ULIRGS, open stars are starbursts). Triangles are BCDs from a
comparison sample. The dotted and solid lines represent forward and bisector 
fits to the data. The thick dashed line shows the trend predicted by the
final star--formation calibrations of Bell 2003 \cite{bell}.
\label{fig:q}}
\end{figure}

Because of the far-infrared/radio (FIR-R) correlation (e.g. \cite{helou1}), 
it is straightforward to show that the radio luminosity 
in evolved luminous starbursts varies only with SFR \cite{condon2}.
The radio continuum emission below about 10\,GHz is dominated by 
non-thermal emission, which is thought to be proportional to the 
supernova rate \nusnr.
Although the mechanism linking \nusnr\ to the non-thermal radio 
luminosity is complex, depending on poorly understood physics of 
cosmic-ray confinement and diffusion operating over $10^7-10^8$\,yr,
radio continuum surveys have been used to constrain co-moving SFR 
locally, and at redshifts of up to z$\sim$1--2 
\cite{cram,mobasher,sullivan,haarsma}. 

Nevertheless, standard conversion of the radio continuum may not
properly estimate the SFR for all galaxies.  Low-luminosity dwarf
galaxies, particularly BCDs, have higher SFRs than would be estimated
by applying canonical relations \cite{klein} and also have globally
flatter radio spectral indices, clearly reflecting a dominant thermal
component \cite{klein,deeg}.  In these galaxies, particularly those of
low metal abundance, the SFR derived assuming the standard mix of
thermal/non-thermal contributions is underestimated by a factor of
five or more \cite{kobulnicky,beck,hunt}. While rare locally (see
\cite{kunth}), low-metallicity low-mass galaxies might be much more
frequent at early times and high redshifts, given the predictions of
the hierarchical merger models (e.g. Baugh et al. (this volume),
\cite{cole1}.  Indeed, such objects may represent the primordial
``building blocks'' -- or ``sub-galaxies'' \cite{rees} -- in
hierarchical scenarios of galaxy formation.  Consequently, at
redshifts $z>3$, it is possible that the canonical scaling of SFR with
radio continuum does not apply.  Such a systematic mis-estimate of SFR
could strongly skew our interpretation of deep radio continuum
surveys.


\subsection{The radio/far-infrared correlation revisited} 

While the ubiquity of the radio/far-infrared (RFIR) correlation is
well documented (e.g. 
\cite{helou1,cox,devereux,wunderlich,price,garrett02,condon2,sadler02}, 
it is also clear that the slope of the
relation depends on the characteristics of the sample.  Samples
dominated by low-luminosity galaxies have steeper than unit slopes
\cite{devereux,price,xu,sadler00} but unit or flatter slopes are
obtained for high-luminosity samples (e.g. \cite{cox,condon2,yun2}).
This is equivalent to saying that the ratio of FIR-to-radio luminosity
$q$ (\cite{helou1}):
$$ q = \log \left[ \frac{{\rm FIR}/(3.75\times10^{12}{\rm Hz})}
{S_{1.4GHz} }\right]$$
(with FIR = $1.26\times10^{-14} [\,2.58
F_{60\mu{\rm m}} + F_{100\mu{\rm m}}\,]$) increases with decreasing
luminosity \cite{bell}, as shown in Figure \ref{fig:q}.


The slope inflection of the FIR-R correlation at
low luminosities probably stems from a number of factors
involving both the radio continuum and the infrared SED.
Moreover, there is some evidence that both the radio continuum
and FIR underestimate the SFR at low luminosities \cite{bell}.


\subsection{Luminosity, metallicity, and age: variations in 
the radio continuum} 

As noted above, thermal emission increasingly dominates the radio 
spectrum of galaxies with low luminosity and low metallicity 
in general, and this is reflected in
their flatter radio spectral indices \cite{klein,klein02,deeg}. 
The same trend is seen in galaxies with very young star-forming 
regions, where non-thermal radio emission is virtually absent 
\cite{roussel}.
The SFR estimated by assuming a purely thermal spectrum is 
roughly ten times higher than inferred from a ``standard''
mainly non-thermal spectrum \cite{condon2}.

Moreover, optically thick HII regions are very likely to be prevalent
in young starbursts \cite{kobulnicky,beck,beck02}.  This means that
instead of a flat thermal spectrum, we would see a rising or
highly-absorbed spectrum at $\sim$\,1\,GHz, which would again alter
the standard assumptions.  Such an effect would also be expected to
vary with redshift.

\begin{figure}[tpb]
\vspace*{5cm}
\includegraphics{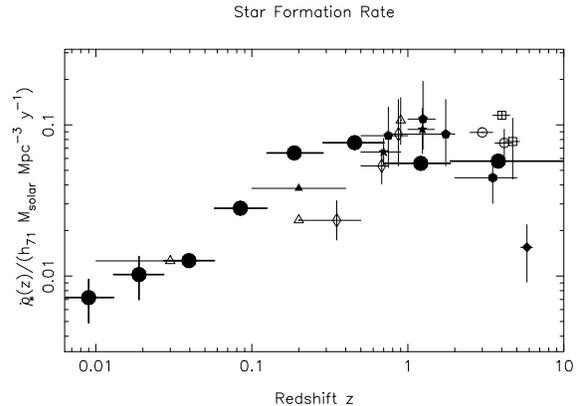}

\caption{The integrated star--formation history of the Universe, from 
\cite{heavens}. The SFR calculated from the ``fossil record'' 
in the Sloan Digital Sky Survey (SDSS) is shown by the eight large filled circles. The other symbols correspond to independent determinations
using instantaneous measurements of the star formation rate at 
various wavelengths.  }
\label{fig:heavens}
\end{figure}

\begin{figure}
\vspace*{8cm}
\includegraphics{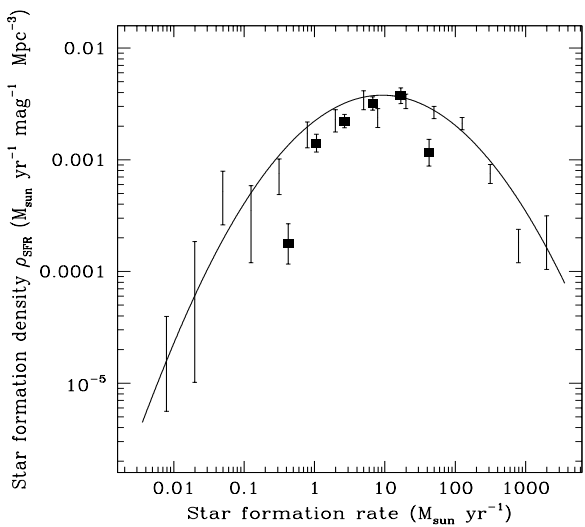}

\caption{Local ($z<0.1$) star--formation density for galaxies with 
star--formation rates between 0.01 and 1000 M$_\odot$\,yr$^{-1}$ 
\cite{sadler00}.  The solid line is derived from a fit to the 
local radio luminosity function for star--forming galaxies, while 
filled squares show values derived from optical measurements of 
the H$\alpha$ emission line \cite{gallego}.  In the local 
universe, the most likely place for a star to form appears to 
be in a galaxy with a star--formation rate around 10\,M$_\odot$ \,yr$^{-1}$.  
The integral under this plot is the zero-point of the 'Madau diagram'  
which plots the integrated star--formation density at different redshifts. }
\label{fig:sfdensity}
\end{figure}

If non-thermal radio emission is suppressed in low-luminosity
galaxies, the disk magnetic field must either be absent or
substantially weaker than in galaxies with higher luminosities (see
\cite{klein,klein02,sadler00}).  If the starbursts are young, it could
also mean that there has not been enough time to set up the usual
$10^8$\,yr diffusion of cosmic ray electrons \cite{helou2}.  If the
radio emission is also compact ($<$200\,pc), such as that in many BCDs
(e.g., \sbs), there can be no $\sim$kpc large-scale diffusion.  This
means that any non-thermal emission must be produced by a different
mechanism (e.g. evolved supernovae expanding in a dense medium
\cite{chevalier,allen}; young supernova remnants; etc.).

In any case, the inescapable implication of the radio spectral
differences as a function of luminosity, metallicity, or age is that
the {\it conversion of radio continuum to SFR may change as a function
of redshift.}  While this is probably not an issue with current
facility surveys which probe only a limited redshift regime, deep SKA
surveys will be able to detect not only the most luminous radio
sources at each redshift, but also adequately sample the statistically
dominant ``normal'' star-forming population out to redshifts of 10 or
so.  Changes in the conversion factors as a function of redshift need
to be considered when mapping cosmic star formation.

%


\subsection{The evolving stellar content of galaxies } 

The integrated star-formation history of the universe (the Madau
diagram; see Fig.~\ref{fig:heavens}) is likely to be accurately mapped
out fairly soon.  This tells us {\it how much}\ star--formation
occurred at a given cosmic epoch, but to understand galaxy evolution
we need to know more than this. In particular, {\it where} and in what
range of environments, did star formation occur at each epoch?  Did
most stars form in galaxies undergoing ``quiescent''star formation, or
in galaxies undergoing violent starbursts, and how does this vary with
redshift?  How important a role do mergers play in triggering star
formation?  In the local universe, star formation occurs in a wide
range of environments, and in galaxies with star--formation rates
ranging over at least five orders of magnitude
(Fig.~\ref{fig:sfdensity}).  There is no reason to suppose that the
high--redshift universe is any less diverse.

Understanding the evolving stellar content of galaxies, and the
chemical enrichment of the interstellar and intergalactic medium which
accompanies this, requires a multi-wavelength approach.  ALMA is sure
to have a big impact, and will be a powerful tracer of star--forming
galaxies over a wide redshift range, detecting
large numbers of normal star--forming galaxies out to z$\sim$3 and Arp
220--like starbursts to z$\sim$10 (e.g. \cite{wootten}). But SKA can 
equally well chart out such luminous star burst galaxies at almost
any redshift and has the great advantage of a much larger field of view
and hence superior survey capabilities. 
Moreover, unlike ALMA, the sensitivity of SKA to star formation does not 
depend on the presence of metals. 
Since the first stars must form in a primordial metal-free environment, 
molecular clouds would not be expected to contain CO or other molecules
beyond molecular hydrogen. 
Hence SKA will provide a
much "cleaner" less biased view of high-redshift star formation. 


%
%

\section{The assembly of galaxies ($0.2 < z < 5$).}

SKA has the unique ability to trace the transformation of HI
gas into stars (and into the enriched hot gas which future X--ray
missions will detect) in a wide range of environments and over a large
fraction of cosmic time.  The fact that HI observations trace the depth
of a galaxy's potential, and so can relate star--formation to halo
mass as well as to luminosity, is also important because it allows
direct tests of galaxy evolution models which deal with mass as well
as light. 

\begin{figure}[htb]
  {\epsfxsize=8.truecm
    \epsfbox[40 70 590 550]{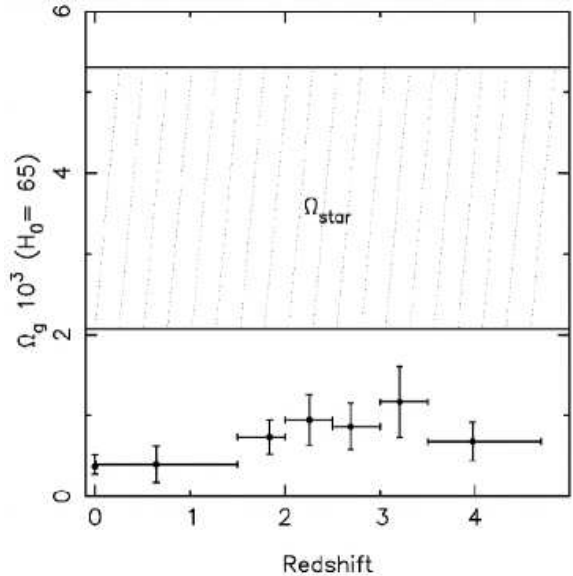}}
  \caption{
Comoving \HI mass density in neutral gas contributed by damped Ly$\alpha$
absorbers plotted as a function of redshift for a cosmology with
$\Omega_M$ = 0.27, and $\Omega_{\Lambda}$ = 0.73 ~\cite{storrielombardie1}.
The region $\Omega_{\rm star}$ is the $\pm 1 \sigma$ range for the mass
density in stars in nearby galaxies \cite{fukugita}. 
The point at z = 0 is the value inferred from 21 cm emission from local
galaxies \cite{zwaan97}.
}
  \label{fig:dlyawithz}
\end{figure}

\begin{figure}[htb]
{\epsfxsize=8.truecm
\epsfbox[40 220 590 700]{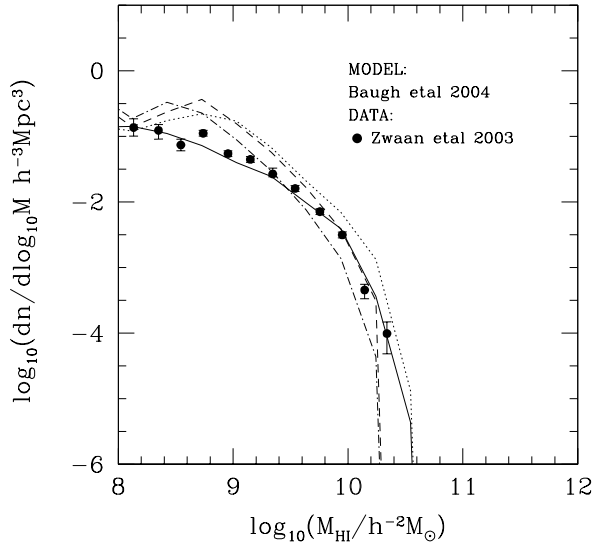}}
\caption{
The evolution of the \HI mass function with redshift in 
the Baugh et al. (this volume) model. 
The points show the recent determination using HIPASS galaxies 
by Zwaan et al.\cite{zwaan3}.
The curves show the model predictions at different redshifts 
(solid: z=0 ; dotted: z=1 ; dashed: z=3, dot-dashed: z=4).
}
\label{fig:mfzzh1a}
\end{figure}

\begin{figure}
{\epsfxsize=8.truecm
\epsfbox[40 70 590 550]{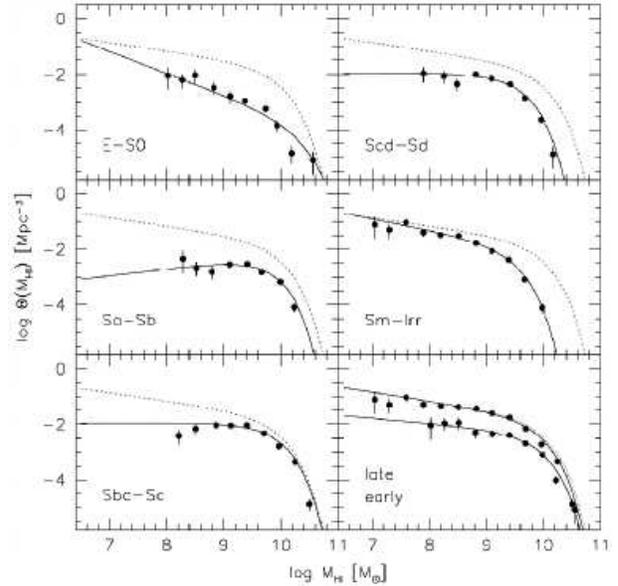}}
\caption{
The \HI mass function for different morphological types. Shown are
Schechter fits (solid lines) and the overall \HI mass function for
the total sample (dashed line). From Zwaan et al.\cite{zwaan3}. 
}
\label{fig:himf-per-type}
\end{figure}

\begin{figure*}[htb]
\includegraphics*[width=6in]{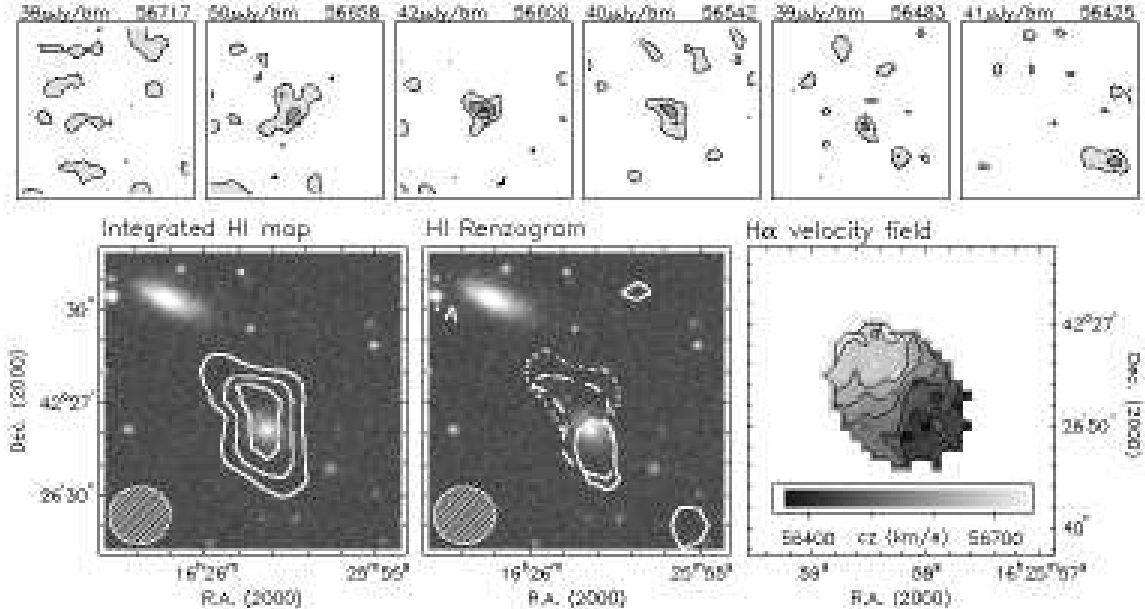}
  \caption{
Atomic Hydrogen detected in Abell 2192 at z=0.1887. {\bf Upper panels:}
individual channel maps from the VLA datacube. The rms noise and
Heliocentric velocity are noted above each panel. {\bf Lower
left:} Total \HI map. The contours coincide with the position of an
inclined barred late-type galaxy. {\bf Lower middle:} contours from the
three channel maps plotted on top of each other; solid: cz=56483 km/s,
dashed: cz=56542 kms, dotted: cz=56600 km/s. {\bf Lower right:} H$\alpha$
velocity field of the optical counterpart obtained with the PMAS IFU
spectrograph at the 3.5m telescope on Calar Alto. There is excellent
correspondence between the \HI and H$\alpha$ recession velocities,
confirming the \HI detection. From Verheijen and Dwarakanath 
(in preparation).
}
  \label{fig:A2192HI.ps}
\end{figure*}

%
%
%
%

\subsection{basic SKA capabilities for \HI emission observations}

Before discussing how SKA will contribute to probing the gas
content of the universe over cosmic time we will briefly describe its
capabilities in terms of sensitivity and resolution to provide a
baseline for \HI imaging projects. \HI line observations typically
require a brightness sensitivity of a few Kelvin, which in practice
limits the angular resolution. For SKA this implies the following:
if one takes as a guideline a $T_b$ sensitivity of $\sim 1$ Kelvin in
a $0.5^{\prime\prime}$ beam at 1.4 GHz (so baselines out to at most a 
few hundred km), 
then one can detect $L_*$ galaxies out to a redshift of $z=1.5$
in a 12 hour integration. Although the current design goals require a 
field of view of one square degree at 1.4 GHz, a much wider field of
view appears technically feasible, and would provide an impressive gain
in survey speed (see also Blake et al. this volume).

Table~\ref{table:hilimits} gives an indication of
the 5$\sigma$ \HI detection limits for a 12 hour integration assuming a
velocity width of 100 km sec$^{-1}$. The calculations assume H$_0$
= 71 km sec$^{-1}$ Mpc$^{-1}$, $\Omega_M$ = 0.27, and
$\Omega_{\Lambda}$ = 0.73. The calculations followed the SKA strawman
design figures: A$_{rm e}$/T$_{\rm sys}$ = 20000, 70\% of the collecting area
is within 100 km and assumed two polarizations. 

\begin{table*}[htb]
\caption{Detection limits for \HI emission with SKA~$^{(1)}$ }
\vspace*{0.15cm}
\label{table:hilimits}
\newcommand{\m}{\hphantom{$-$}}
\newcommand{\cc}[1]{\multicolumn{1}{c}{#1}}
\renewcommand{\arraystretch}{1.2} 
\begin{tabular}{@{}|ccccccccc|} 
\hline
$z$&Freq.&T$_{\rm sys}$~$^{(2)}$&Angular~$^{(3)}$&Linear&SB&Luminosity&Lookback&\HI mass~$^{(4)}$ \\ 
        & & &Resol.& Resol. &Dimming&Distance&Time&limit \\   
        &(MHz)&(K) &(arcsec)&(kpc)    &(mag)&(Gpc)&(Gyr)& (\Msun) \\
\hline
0.2  &1183.67&50.4&0.52 & 1.7  & 0.796 & 0.972 & 2.41 &$6.1 \times 10^8$    \\
0.5  & 946.94&51.4&0.65 & 4.0  & 1.486 & 2.825 & 5.02 &$8.7 \times 10^8$    \\
1.0  & 710.20&53.8&0.87 & 7.0  & 3.026 & 6.640 & 7.73 &$2.7 \times 10^9$    \\
1.5  & 568.16&57.5&1.09 & 9.3  & 4.000 & 11.02 & 9.32 &$7.2 \times 10^{9}$ \\
2.0  & 473.47&62.7&1.31 & 11.1 & 4.796 & 15.75 & 10.32 &$1.5 \times 10^{10}$ \\
2.5  & 405.83&69.6&1.52 & 12.5 & 5.469 & 20.72 & 11.00 &$2.6 \times 10^{10}$ \\
3.0  & 355.10&78.3&1.74 & 13.6 & 6.052 & 25.87 & 11.48 &$4.3 \times 10^{10}$ \\
3.5  & 315.64&89.3&1.96 & 14.6 & 6.566 & 31.15 & 11.83 &$6.7 \times 10^{10}$ \\
\hline 
\end{tabular}\\[2pt] 

\vspace*{0.1cm}
\hspace*{0.2cm}$^1$ ~assuming t = 12 hours, A$_{\rm e}$/T$_{\rm sys}$=20000, 
	   2 polarizations and 70\% of A$_{\rm e}$ within 100 km)\\
\hspace*{0.2cm}$^2$ ~including a contribution from Galactic foreground emission assuming T$_{\rm Gal}(f_{\rm MHz})$ = 20$(\frac{408}{f_{\rm MHz}})^{2.7}$ K \\
\hspace*{0.2cm}$^3$ ~fixed array geometry assumed so that resolution scales with wavelength\\
\hspace*{0.2cm}$^4$ ~assuming 5 rms and 100 km sec$^{-1}$ profile width. At z=0.2 and z=0.5
the galaxies are assumed resolved 
\hspace*{0.45cm} so here the flux has been added over 8.5 and 1.5 beams 
respectively.
\end{table*}

It is clear from Table~\ref{table:hilimits} that it is possible to detect
large galaxies in \HI emission beyond a redshift of $z = 1$ fairly
easily, though with limited resolution.  For example a galaxy like the
Milky Way can be detected out to a redshift of about 1 in a 12 hour
integration. Its companions, the Magellanic Clouds, can still be seen
out to redshifts of a few tenth. Large spiral galaxies such as M~101
can be probed much farther, to $z \approx 2.5$. 


\subsection{The evolving gas content of galaxies}

Various lines of evidence suggest that there is strong evolution
between $z = 1$ and $z = 3$. This is shown not only by the latest
estimates of the change in comoving star formation density with
redshift \cite{heavens} but also by the most recent estimates of the
evolution in gas density with redshift from damped Ly$\alpha$ studies
 \cite{storrielombardie1,giavalisco1}.

Direct measurements of the evolution of the \HI in galaxies out to
redshifts of 1 and beyond are not yet available, because present day
instruments lack the sensitivity and resolution to directly measure
the \HI in galaxies at these redshifts. The largest redshift \HI detection
to date is a deep VLA observation of the cluster Abell 2192 at $z = 0.1887$
\cite{verheijen3} which detected \HI in a late type galaxy
at this redshift (see Fig.~\ref{fig:A2192HI.ps} ). Scaling this observation to 
the capabilities of SKA one expects to get similar results for galaxies 
at distances beyond $z = 1$.   

Damped Ly$\alpha$ studies
~\cite{storrielombardie1} indicate that the comoving \HI mass density
is a few times the present value beyond $z = 1$ and out to $z = 3$ as
illustrated in Fig.~\ref{fig:dlyawithz}. The evolution in comoving \HI
mass density as shown in Fig.~\ref{fig:dlyawithz} is still dependent on
assumptions. A recent study of the column density distribution
function of \HI from observations of nearby galaxies by Zwaan et al.
~\cite{zwaan3} suggests that
there is not as strong an evolution in comoving \HI mass density as
suggested in Fig.~\ref{fig:dlyawithz}. SKA, however,
will be able to measure the \HI emission in galaxies back to redshifts
of $z \approx 3$ directly and will revolutionize this area of
research.

There are various ways in which the SKA will contribute significantly
to determining the way in which galaxies assemble over cosmic time.
Since the evolution of comoving \HI gas density as described above has
not been directly measured, but has been derived from \HI column
density measurements of Damped Ly$\alpha$ absorbers it is crucial to
be able to image the entire \HI disks associated with these Damped
Ly$\alpha$ absorbers. If the associated \HI masses are indeed of the
order of M$^*_{\mHI}$ = $6.2 \times 10^9$ M$_{\odot}$
\cite{zwaan1,zwaan2,zwaan3} then SKA can image these disks out to
$z = 1.4$ in integration times of 12 hours. Such observations will be
crucial for linking the existing column density measurements to the
local \HI density from observations of the emission.

Another way to probe the evolution of galaxies over cosmic time is to
determine variations in the \HI Mass Function as a function of
environment (local galaxy density) and redshift. From recent studies
(see Fig.~\ref{fig:himf-per-type} from Zwaan et al. \cite{zwaan3}) it
is already apparent that the \HI mass function depends on Hubble type. 
The \HI mass function also appears to depend on environment. The slope
of the overall HI mass funtion is steeper ($\sim - 1.3$ \cite{zwaan3})
than in the Local Group \cite{zwaan4}~and Ursa Major cluster 
\cite{verheijen4} where the slope is $\sim -1.0$.
So by measuring the \HI mass function over a range
of redshifts and in different environments one can probe the effects
of galaxy evolution. The models of Baugh et al. (this volume) also
indicate changes in the \HI mass function (see Fig.~\ref{fig:mfzzh1a})
which SKA will be able to determine.

In order to do this reliably, sufficient numbers of galaxies over a
range of \HI masses need to be detectable. A deep \HI survey using of
order 30 days of integration time will be able to detect L$_*$
galaxies (which typically have \HI masses of $6.2 \times 10^9$ M$_{\odot}$
assuming H$_0$ = 71 km/s/Mpc) out to redshifts of $z
\approx 2.7$, i.e.  beyond the redshift where the universe shows
considerable evolution.  Using the \HI mass function of Zwaan et
al.(2003)~\cite{zwaan2} and assuming no evolution of the \HI
properties of galaxies with redshift one can calculate how many
galaxies one would expect to detect per redshift interval in a one
square degree field of view.  Table~\ref{table:deephisurveyspecs}
gives a brief summary. These calculations were performed for a
$\Omega_M=0.27$ and $\Omega_\Lambda=0.73$ cosmology.
In such long integrations the emission
associated with Damped Ly$\alpha$ systems in the field of view 
can be detected out to $z \approx 2.7$ 
(assuming they are $\approx$ M$^*_{\rm{HI}}$ systems).

\begin{table*}[htb]
\caption{Detectable \HI Masses as a function of redshift and number of
detected galaxies per \newline square degree in 360 hour and 1000 hour
integrations} \vspace*{0.15cm}
\label{table:deephisurveyspecs}
\newcommand{\m}{\hphantom{$-$}}
\newcommand{\cc}[1]{\multicolumn{1}{c}{#1}}
\begin{tabular}{@{}|cc|cc|cc|}
\hline
Redshift   & Look Back Time & \HI Mass Limit~$^{(1)}$ & Number 
&\HI mass limit~$^{(1)}$ & Number \ \\
\          &  (Gyr)         & ( M$_\odot$)  & of detections
& ( M$_\odot$)& of detections\  \\
& & t = 360 h & & t = 1000 h & \ \\
\hline
0.5 -- 1.0 &  5.0 -- 7.7 &$5.0 \times 10^8$   &$1.6 \times 10^6$&$3.0 \times 10^{8} $&$2.4 \times 10^6$  \\ 
1.0 -- 1.5 &  7.7 -- 9.3 &$1.3 \times 10^9$   &$1.8 \times 10^6$&$7.9 \times 10^{8} $&$2.7 \times 10^6$  \\ 
1.5 -- 2.0 &  9.3 -- 10.3&$2.7 \times 10^9$   &$1.9 \times 10^6$&$1.6 \times 10^{9} $&$2.7 \times 10^6$  \\ 
2.0 -- 2.5 & 10.3 -- 11.0&$4.8 \times 10^9$   &$1.7 \times 10^6$&$2.9 \times 10^{9} $&$2.5 \times 10^6$  \\ 
2.5 -- 3.0 & 11.0 -- 11.5&$7.8 \times 10^{9}$ &$1.6 \times 10^6$&$4.7 \times 10^{9} $&$2.3 \times 10^6$  \\ 
3.0 -- 3.5 & 11.5 -- 11.8&$1.2 \times 10^{10}$&$1.4 \times 10^6$&$7.3 \times 10^{9} $&$2.0 \times 10^6$  \\ 
3.5 -- 4.0 & 11.8 -- 12.1&$1.8 \times 10^{10}$&$1.2 \times 10^6$&$1.1 \times 10^{10}$&$1.8 \times 10^6$  \\ 
4.0 -- 4.5 & 12.1 -- 12.3&$2.7 \times 10^{10}$&$1.0 \times 10^6$&$1.6 \times 10^{10}$&$1.5 \times 10^6$  \\ 
\hline
\end{tabular}\\[2pt] 

\vspace*{0.1cm}
\hspace*{0.2cm}$^{1}$ ~calculated for the larger $z$ in each interval. Other 
assumtions as in table 1.
\end{table*}

Out to $z \approx 0.3$ one will be able to resolve a fair fraction of
the galaxies so one can obtain rotation curves, mass distributions, and
gas fractions for more than $\approx 10^4$ galaxies between now and 3 Gyr
ago.  One then is in a unique position to trace the evolution of the
ISM in galaxies over a substantial fraction of the age of the
universe, from the era of strongest evolution and star formation
activity until the present. In addition one would learn whether and
how the evolution depends on the dark matter content and environment.

The hierarchical galaxy formation simulations suggest that most of the
evolution in the HI mass function is in the mass range M$_{\HI}$ = 4
$\times$ 10$^8$ -- 4 $\times$ 10$^9$ \Msun and within the last seven
billion years.  Fig.~\ref{fig:mfzzh1a} shows that at $z = 1$ one
expects $\sim 3$ times more galaxies in this mass range than at $z
= 0$.  This figure also illustrates that there is very little
evolution in the high mass ( M$_{\rm{\HI}}$ $> 4 \times$ 10$^9$ \Msun) end
of the \HI mass function.  This is the range where the SKA is
especially sensitive and can probe easily to at least $z = 1$, so that
this prediction can be verified.

If, on the other hand, the suggestion that the gas accretes along
filaments and is not heated to the virial temperature 
~\cite{binney1,katz1,murali1} is correct, then a deep SKA survey 
should be able to
see the filamentary cool gas reservoirs out to $z = 3$ since until 
$z \approx 1$ there is as much gas in filaments as there is in galaxies.
Since in this picture the size of the filaments that feed the galaxies
grow from galaxy size at high redshift, to hundreds of kpc size structures
feeding entire groups at low redshifts, deep \HI surveys should be able 
to verify this idea. If on the other hand mergers of predominantly 
small systems are the main process one expects to find a very different
distribution of \HI structures with redshift. 

In addition, as detailed in the previous section, one can use the
continuum emission to probe the star formation rates of the detected
galaxies, independent of the effects of extinction using the radio
continuum-FIR correlation
~\cite{helou1,helou2,condon1,condon2,sauvage1} and test the model
predictions.  In addition, this information can be used to link the
star formation rates to the \HI contents of galaxies as a function of
redshift and environment. The global star formation rates can be
compared with the semi-analytic models and the cool gas accretion
models in order to test the predictions. That environment plays an
important role is suggested by the analysis of Sloan Digital Sky
Survey (SDSS) data \cite{kauffmann,kauffmann2,hopkins1,hogg1} which
shows a dependence of galaxy mass and star formation rate on
environment.

\section{Evolution now ($z < 0.2$).}

\begin{figure}
  {\epsfxsize=10.5truecm
    \epsfbox[80 50 630 530]{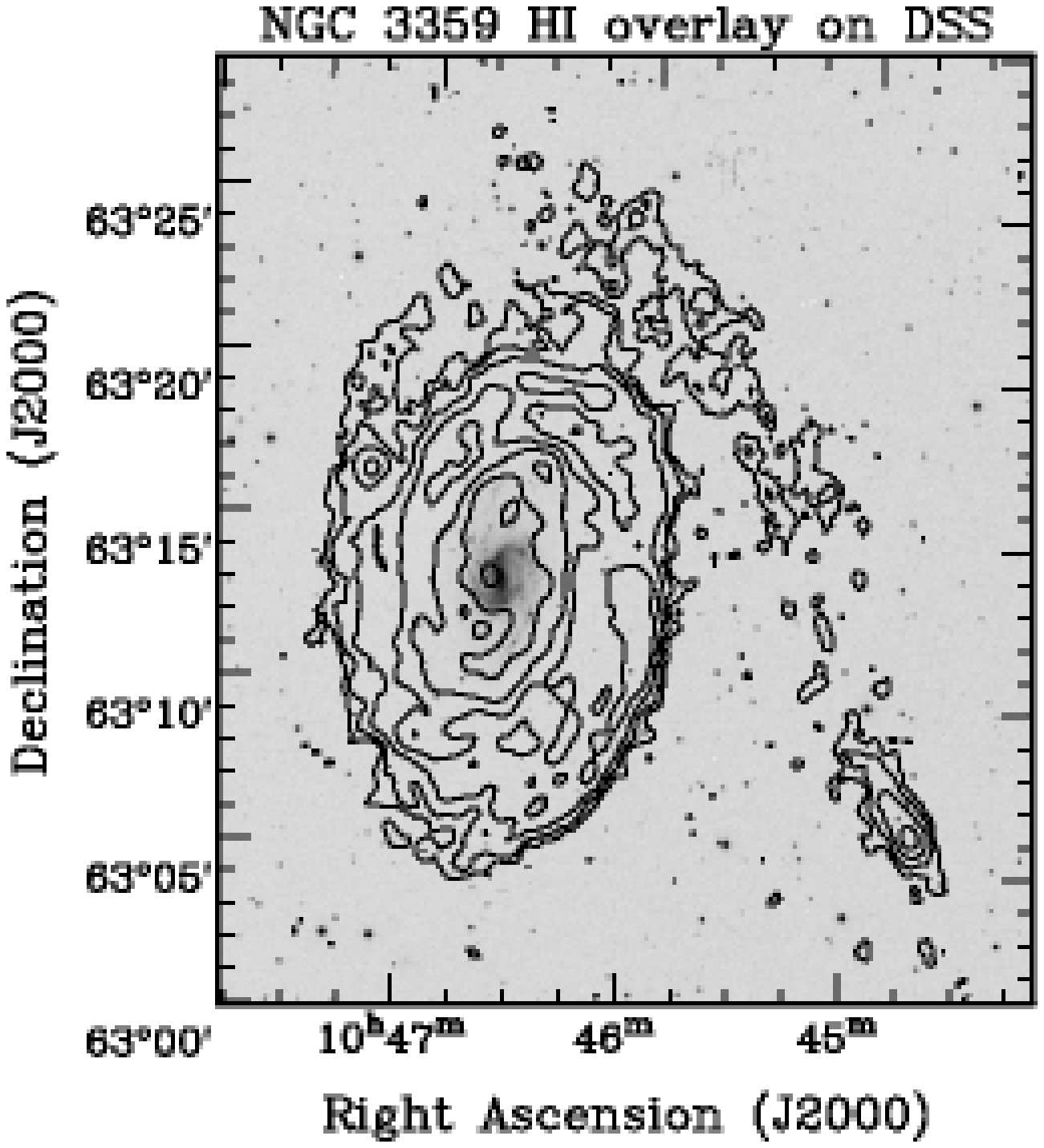}}
  \caption{\HI distribution of NGC~3359 at a resolution of 
30$^{\prime\prime}$ superposed on the digital sky survey 
image (left panel). Contours are 0.1, 0.2, 0.4, 0.8, 1.6, 3.0, and 5.0 
$\times 10^{21}$ cm$^{-2}$~\cite{vdhulst1}}
  \label{fig:n3359hi}
\end{figure}

The process of merging and accretion continues to the present
time. Ample evidence is already provided from existing observations,
both optical and \HI imaging. Zaritsky \cite{zaritski1} and Zaritsky
\& Rix \cite{zaritski2} studied several tens of galaxies and ascribed
the coincidence of evidence for recent star formation from stellar
population studies with measurable asymmetries in the galaxy shapes,
to recent accretion events. Kinematic lopsidedness, observed in the 2
dimensional \HI velocity fields of galaxies \cite{verheijen1,swaters1}
has also been considered as a result of recent minor
mergers. Furthermore there are at least twenty examples of
galaxies which in \HI show either signs of interactions and/or have
small companions \cite{sancisi1,sancisi2,vdhulst1}.  This suggests
that galaxies often reside in an environment where material for accretion
is available. An example is given in Fig.~\ref{fig:n3359hi} which
shows a new \HI image of the galaxy NGC~3359 which is accreting gas
from a faint \HI companion.

Other clear evidence for ongoing minor merger activity comes from the
discovery of the Sagittarius dwarf galaxy \cite{ibata1} and of
substructure in the halo of the Local Group galaxy M~31
\cite{ibata2,ferguson1,mcconnachie1}.

\begin{figure*}[htb]
\includegraphics*[width=6in]{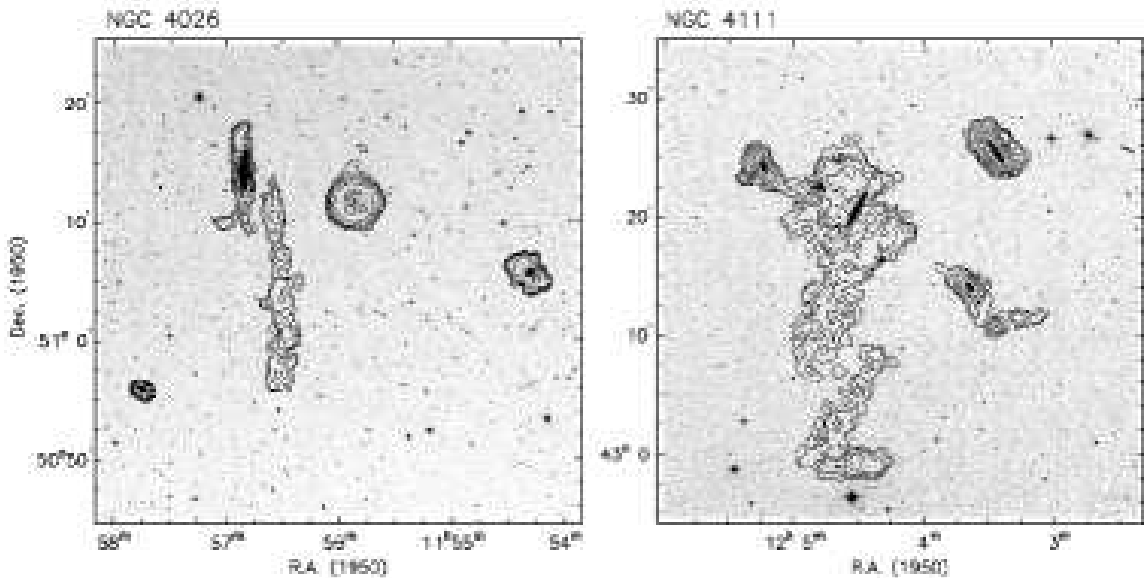}
  \caption{
Extended \HI filaments observed near the brightest lenticular galaxies in
the Ursa Major cluster. These lenticulars are located in small
sub-groups within the Ursa Major volume. and the \HI morphology indicates 
that tidal interaction between the galaxies is taking place. 
From Verheijen and Zwaan \cite{verheijen2}
}
  \label{fig:N4026N4111_xv}
\end{figure*}

\begin{figure}[tpb]
  {\epsfxsize=8.truecm
    \epsfbox[40 170 590 650]{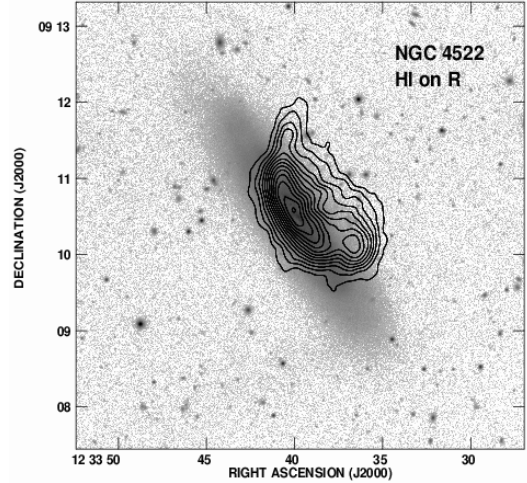}}
  \caption{
NGC~4522 in the Virgo cluster from Kenney et al. \cite{kenney1}. 
The \HI contours are displaced from the optical disk and clearly 
show the sweeping effect of ram-pressure stripping.
}
  \label{fig:kenney.fig3}
\end{figure}

The \HI Rogues Gallery of Galaxies \cite{hibbard1},
~{\it www.nrao.edu/astrores/HIrogues/}) gives a glimpse of the role of
gas in galaxy evolution at $z = 0$.  Ideally one would want to make
complementary Rogues Galleries at $z = 0.5$, $z = 1$, $z = 1.5$, and
$z = 2$ to study the process of accretion in detail. Inspection of
the Rogues Gallery clearly demonstrates the importance of using the
\HI for probing galaxy evolution. The prospect of obtaining similar
information to either higher depth in sensitivity or to much larger
redshifts is truly exciting as radio observations are the only means
to probe these aspects of galaxy evolution. There simply is no other
way to trace the neutral atomic gas around galaxies. A very good
illustration is the extended \HI around SO galaxies in the Ursa Major
cluster shown in Fig.~\ref{fig:N4026N4111_xv}. SO galaxies are thought
to be the result of galaxy transformation by gas removal (stripping
and/or tidal effects) but in these objects this is not so obvious. A
second example of the diagnostic power of \HI imaging is NGC~4522 in
the Virgo cluster, where the \HI morphology presents overwhelming
evidence that stripping by ram pressure is taking place. (see
Fig.~\ref{fig:kenney.fig3})

\begin{figure}
  {\epsfxsize=8.truecm
    \epsfbox[40 70 590 550]{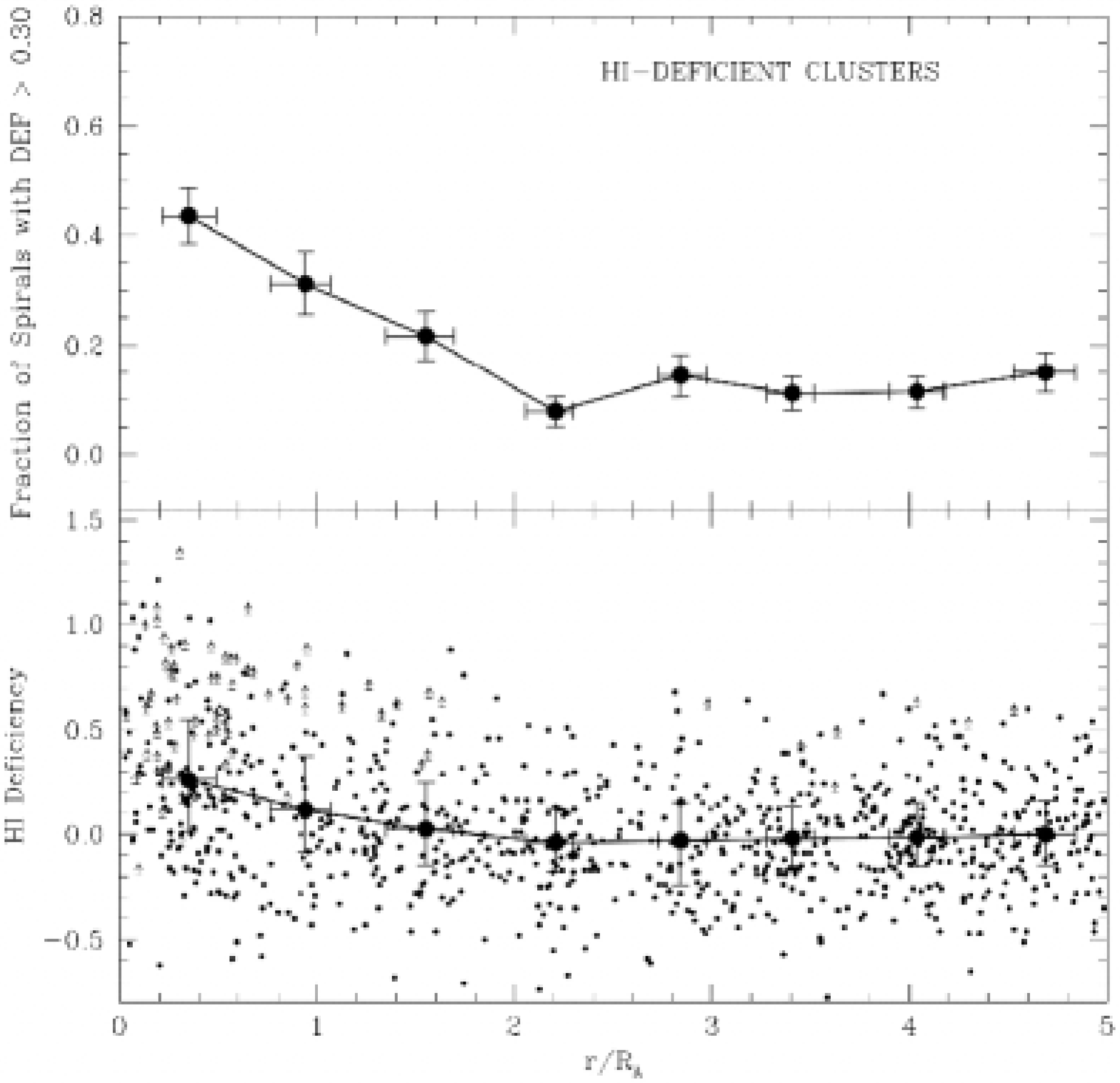}}
  \caption{Top: \HI deficient fraction in bins of projected radius 
from the cluster center for the superposition of all the \HI deficient 
clusters. Bottom: \HI deficiency versus projected radius from the cluster 
center. Small dots show   the radial variation of \HI deficiency for 
individual galaxies, while the arrows indentify non detections plotted
at their estimated lower limit. Large dots are the medians of the 
binned number distribution. From Solanes et al. \cite{solanes1}.
}
  \label{fig:hideficiency}
\end{figure}

\begin{figure}
  {\epsfxsize=8.truecm
    \epsfbox[40 50 590 530]{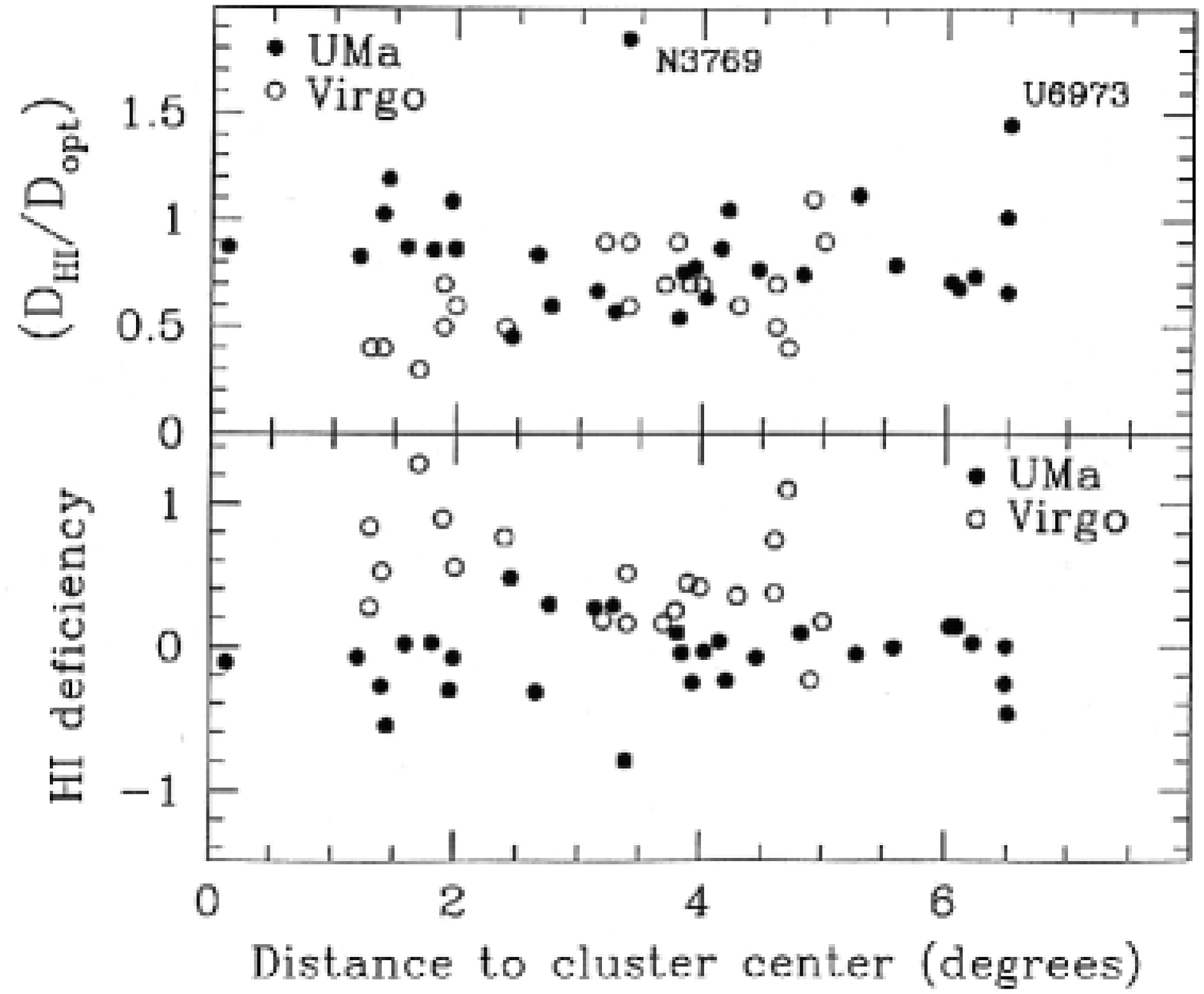}}
  \caption{
A comparison of \HI properties of galaxies in the Ursa Major and Virgo
clusters as a function of projected distance to the cluster center.
Upper panel: ratio of \HI to optical diameters. 
Lower panel: overall \HI deficiency as a function projected distance.
}
  \label{fig:UMaVirgoHIdef}
\end{figure}

That indeed the environment plays an important role in determining the
\HI properties (and hence probably the evolution) of galaxies has
become clear from different lines of evidence (for a recent review see
van Gorkom \cite{vangorkom2}): the finding that galaxies in clusters
out to 2 Abell radii are deficient in \HI as compared to galaxies of
similar type in the field \cite{solanes1}, see also
Fig.~\ref{fig:hideficiency} ) and the truncated and sometimes displaced
\HI distributions of galaxies in rich environments (resulting from
ram pressure stripping and tidal effects). Additional evidence for
strong environmental effects is shown in Fig.~\ref{fig:UMaVirgoHIdef}
from Verheijen et al. ~\cite{verheijen2}. In the denser environments
of the core of the Virgo cluster the \HI deficiencies and \HI to
optical sizes are notably different (respectively larger and smaller)
than in the less dense environment of the Ursa Major cluster and the
field.

The velocity structure and spatial distribution of \HI detected
galaxies near clusters suggests that infall is still going on and that
clusters are still capturing galaxies which fall in along large scale
structure filaments. Evidence such as the \HI structures around S0
galaxies and the stripped gas in NGC~4522 (Fig.~\ref{fig:N4026N4111_xv} 
and Fig.~\ref{fig:kenney.fig3}) are crucial in pinning
down these processes and SKA will be able to see such structure over a
considerable range of redshift (i.e. evolution).

A shallow survey covering 100 square degrees in two months (assuming
100 km sec$^{-1}$ profile width, a 5$\sigma$ detection limit, a field
of view of one square degree at 1.4 GHz
\footnote{if the SKA has a field of view of 50 square degrees rather than 
the one square degree of the SKA strawman design, such a survey could be 
carried out in one day}, and an instantaneous bandwidth of 500 MHz)
will be able to see NGC~3359--like structures to $z=0.15$ opening
the exciting prospect of being able to probe the environment of
several million galaxies and pin down the merger/accretion
characteristics at the present time over a range of galaxy
densities. The resolution and sensitivity of such a survey will be
comparable to that of the images shown in Fig.~\ref{fig:n3359hi} and
Fig.~\ref{fig:N4026N4111_xv}.  The volume covered by such a survey is
$\sim$ 15,000 Mpc$^3$ and will detect $\sim$ half a million
galaxies, providing a detailed \HI image, not only of the galaxies but
also of their \HI environments.  Only with such data can the evolution
of galaxies, in particular the mechanisms for converting gas into stars
in and around galaxies be studied.

\end{document}